\def\cm2{cm$^{-2}$}

\def\c2{C~{\sc ii}}
\def\c4{C~{\sc iv}}
\def\fe2{Fe~{\sc ii}}
\def\fe3{Fe~{\sc iii}}
\def\mg1{Mg~{\sc i}}
\def\mg2{Mg~{\sc ii}}
\def\si2{Si~{\sc ii}}
\def\si4{Si~{\sc iv}}
\def\al2{Al~{\sc ii}}
\def\al3{Al~{\sc iii}}
\def\o1{O~{\sc i}}
\def\n1{N~{\sc i}}
\def\h1{H~{\sc i}}

\def\approxlt{\mathrel{\spose{\lower 3pt\hbox{$\sim$}}
        \raise 2.0pt\hbox{$<$}}}
\def\approxgt{\mathrel{\spose{\lower 3pt\hbox{$\sim$}}
        \raise 2.0pt\hbox{$>$}}}

\documentclass[article]{gwp80}
\usepackage{graphicx}
\usepackage{amssymb}
\tabletypesize{\normalsize}  % AMcW

\def\plotone#1{\centering \leavevmode
\includegraphics[width=.95\columnwidth]{#1}}

\def\plotone#1{\centering \leavevmode
\includegraphics[width=.95\columnwidth]{#1}}

\shortauthors{Kolenberg}
\shorttitle{Blazhko Stars}

\begin{document}
\large    %AMcW  The conference proceedings will employ large size print
\pagenumbering{arabic}
\setcounter{page}{100}

\title{Peculiarities of Blazhko Stars: New Insights}

%
% Here is an example of how to include the author names and affilitations
%
\author{{\noindent Katrien Kolenberg{$^{\rm 1,2,3}$} \\
%for KASC RR Lyrae working group\\
\\
{\it (1) Harvard-Smithsonian Center for Astrophysics, 60 Garden Street, Cambridge MA 02138, USA\\
(2) Instituut voor Sterrenkunde, K.U.Leuven, Celestijnenlaan 200D, 3001 Leuven, Belgium\\
(3) Institut f\"ur Astronomie, T\"urkenschanzstrasse 17, A-1180 Vienna, Austria\\} 
}
}

%
% And here is how to add the e-mail addresses
%
\email{(2) kkolenbe@cfa.harvard.edu}

% If you really need to add an alternate institution, then update and uncomment the following line.
% It's not very pretty though
%\altaffiltext{}{(2) Harvard}

\begin{abstract}
With increasingly accurate data on RR Lyrae stars we find that the Blazhko effect may be a rule rather than an exception. 
However, we still do not know what is the cause of this mysterious amplitude and phase modulation.
In my talk, I intended to give a glimpse of the properties of Blazhko stars, presenting recent findings concerning what makes Blazhko stars different from their non-modulated counterparts.
Recent observations of Blazhko stars obtained from space
%, with the {\it CoRoT} and {\it Kepler} telescopes, 
give some important clues that deserve further exploration.
\end{abstract}

\section{A century-old mystery}

In 1907, Sergei Nicolaevich Blazhko reported on the periodic variations in the timing of maximum light for the star RW Dra (Blazhko 1907).  Soon after that, it was realized that also other, similar stars, showed modulations of their light curve shape over time scales of weeks or even months.  In 1916 Harlow Shapley reported "on the changes in the spectrum, period and light curve of the Cepheid variable RR Lyrae", the star that later became the eponym of a new, distinct, class of variables.
Szeidl (1988) reported an occurrence rate of 20-30\%  for Galactic RRab stars and only a few percent for RRc stars.  In the LMC this occurrence rate appears to be somewhat lower (12 \% RRab and 4\% RRc, see Alcock et al. 2003, 2000 respectively), though high-precison surveys may slightly change these numbers.
The most recent surveys, both from the ground (Jurcsik et al. 2010) and from space (Szab\'o et al. 2010a, Benk\H{o} et al. 2010), seem to indicate that close to 50\% of the fundamental mode RR Lyrae pulsators show the Blazhko effect.  Typical light curves of a non-Blazhko and a Blazhko variable obtained with the {\it Kepler} space telescope are shown in Figure\,1. 

More than a hundred years after its discovery, and with increasingly higher occurrence rates for the Blazhko effect, we still are at loss for an explanation.

\begin{figure*}
\centering
\plotone{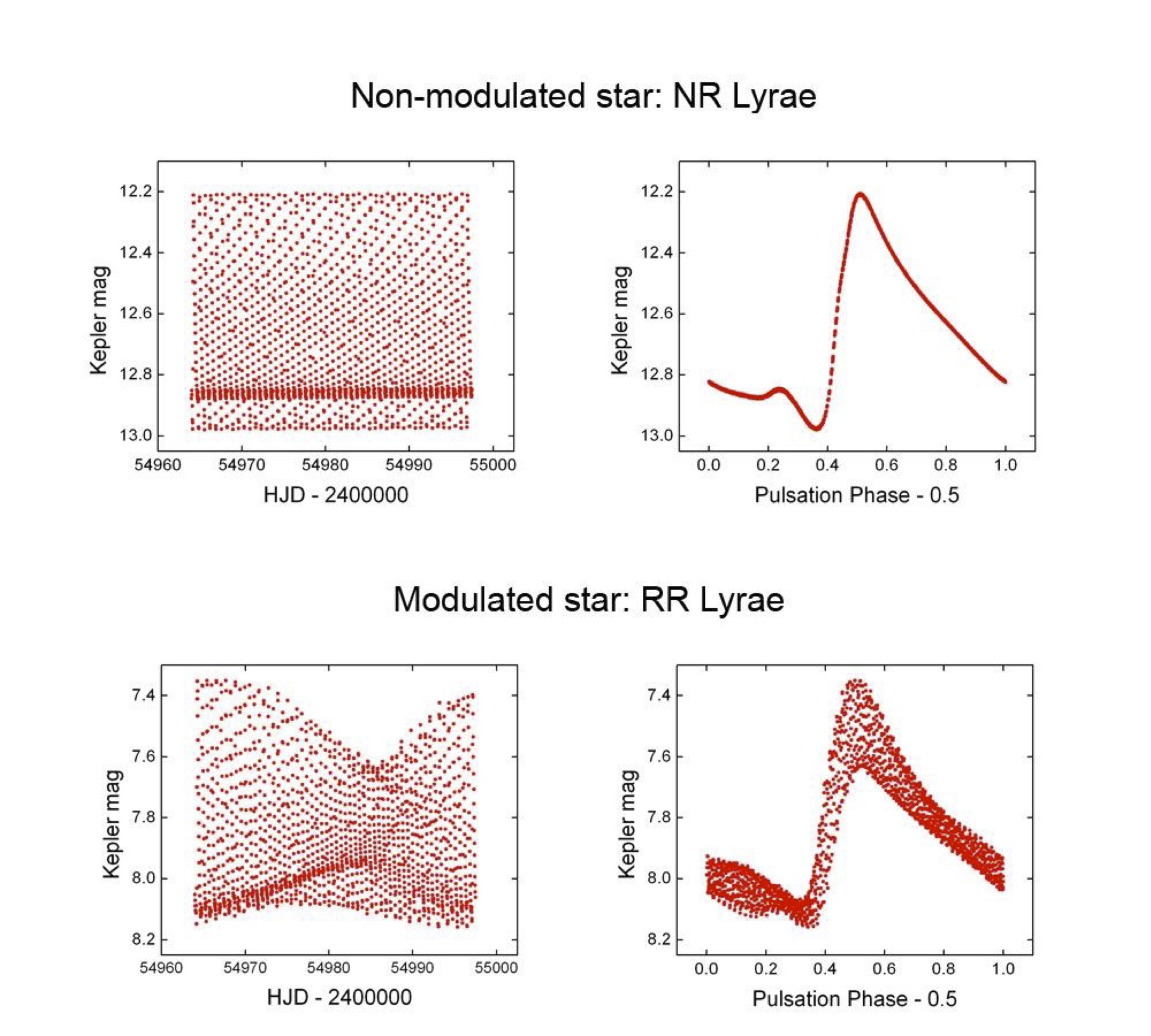}
%\plotone{Kolenberg_Figure1.eps}
\vskip0pt
\caption{The upper panel shows direct and folded Q1 light curves for the non-modulated RR Lyr star NR Lyr, while the lower panel illustrates the rapid changes of the light curve of RR Lyr itself (the Blazhko effect), characteristic of many RR Lyrae stars monitored with Kepler.     }
\label{o1039}
\end{figure*}

\subsection{Why do some stars do it and others don't?}

\begin{figure*}
\centering
%\includegraphics[width=12cm]{mgsio2.eps}
%\plotone{Kolenberg_Figure2.eps}
\plotone{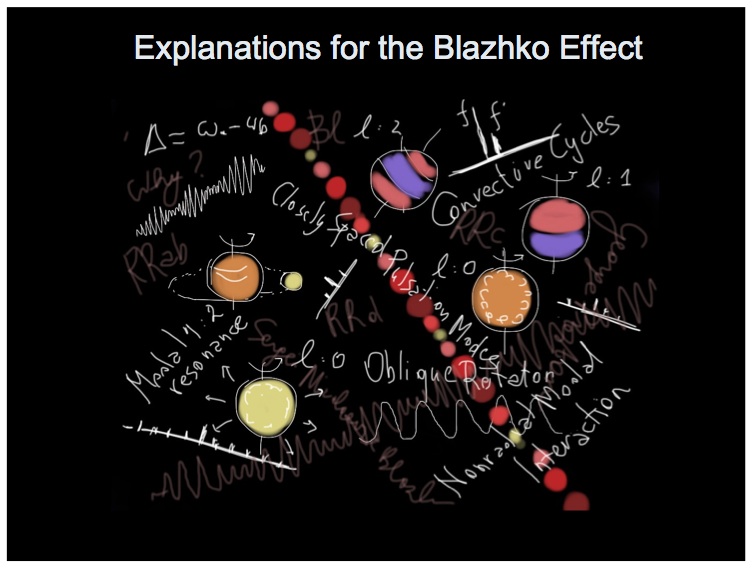}
\vskip0pt
\caption{More than a century after the discovery of the Blazhko effect, we still are at loss for a definitive explanation.  But we are narrowing down the possibilities.}
\label{o1039}
\end{figure*}

A plethora of models for the Blazhko effect have been suggested over the course of the past century (Figure\,2). 

Since about 1995 and until just a few years ago, two models for the Blazhko effect were mostly quoted in the literature: the resonance model (Van Hoolst et al. 1998, Dziembowski \& Mizerski 2004, and references therein) and the magnetic model (Shibahashi \& Takata 1995, Shibahashi 2000, and references therein). Both of them rely on the excitation of nonradial pulsation mode components in the modulated star. Both models also state a connection between the modulation period and the rotation period, and provide predictions for the appearance of the Fourier spectra of modulated stars.

\begin{figure*}
\centering
%\includegraphics[width=12cm]{mgsio2.eps}
%\plotone{Kolenberg_Figure3.eps}
\plotone{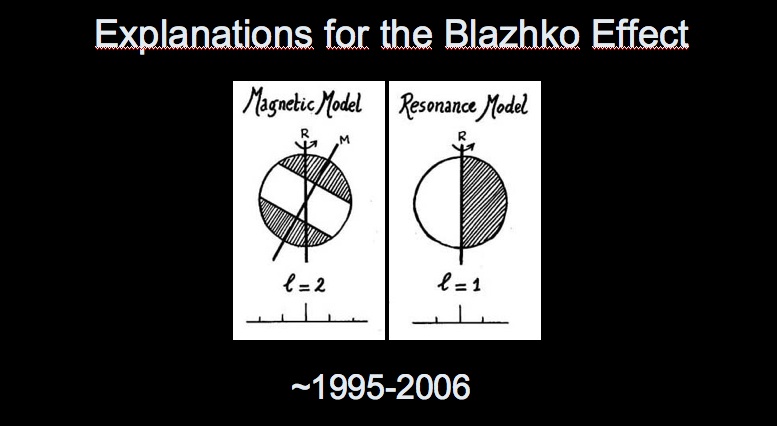}
\vskip0pt
\caption{The mostly quoted models for explaining the Blazhko effect for more than a decade.}
\label{o1039}
\end{figure*}

\vspace{2mm}

The variety of observed behavior in Blazhko stars did not agree well with the predictions by the prevalent models.  Such findings were, among others, the observation of multiple modulation periods (LaCluyz\'e et al. 2004, S\'odor et al. 2011), changing modulation periods (LaCluyz\'e et al. 2004, Kolenberg et al. 2006), and complex multiplet structures beyond the triplets and quintuplets predicted by the resonance and magnetic models (Jurcsik et al. 2008). In the light of this, an alternative idea was proposed by Stothers (2006, 2010).  In this scenario, the amplitude and phase (period) modulation is caused by the cyclical weakening and strengthening of the convective turbulence in the star. The variations in convective turbulence can be caused by a transient magnetic field in the star.  The presence of such a field, however, would be very hard to demonstrate (see Kolenberg \& Bagnulo 2009).  The stochastic nature of the scenario proposed by Stothers (2006) makes it an attractive idea, as it we do not see clockwork regularity in the Blazhko cycles of several well-studied Blazhko stars (e.g., LaCluyz\'e et al. 2004, Kolenberg et al. 2011, Guggenberger et al. 2011). Also, the variations of the mean parameters of the star, as follow from Jurcsik et al. (2009) measurements  of the star MW Lyr,  are a consequence of the modulation of turbulent convection in this model.  Moreover, the Stothers (2006) scenario does not require any basic change in our current understanding of RRab and RRc stars as being purely fundamental-mode and first-overtone radial pulsators. Therefore, it has recently gained popularity in the literature.  A critical discussion of the Stothers (2006) idea was given by Kov\'acs (2009) and in Section\,4 I will briefly discuss a closer examination of the model.

\section{New developments}

In this section I want to point out a few relatively recent studies that aimed at establishing what makes Blazhko stars different (hereby apologizing for incompleteness).

When looking at the amplitude (ratios) of the photometric and radial velocity changes of Blazhko RRab stars in their different Blazhko phases, it can be seen that a normal amplitude (ratio) occurs near the phase of maximum amplitude. 
Therefore, the Blazhko effect seems to quench the stellar pulsation, and at maximum amplitude the light curve might be like that of a non-modulated star. Jurcsik et al. (2002) checked the Fourier parameters of maximum amplitude light curves with different methods and showed that none of these light curves can be regarded as a normal RR Lyrae type light curve, and thus, that Blazhko stars are ``always distorted''.

Using the available data on Blazhko stars, Jurcsik et al. (2005a) found that the possible maximum value of the modulation frequency depends on the pulsation frequency. Short period variables ($P<0.4$ d) can have modulation period as short as a few days, while longer period variables ($P>0.6$ d) always exhibit modulation with $P_{\rm mod}>20$ d. They interpreted this tendency with the equality of the modulation period with the surface rotation period.  Changing (and multiple) modulation periods clash in this framework with the idea of a stable (and constant/single) rotation period. In addition, the possible largest value of the modulation amplitude, defined as the sum of the Fourier amplitudes of the first four modulation frequency components, appears to increase towards shorter period variables (Jurcsik et al. 2005b).  Such observed systematics help in constraining the mechanism causing Blazhko modulation.

S\'odor et al. (2009) devised an inverse photometric Baade-Wesselink method for determining physical parameters of RRab variables exclusively from multicolour light curves.  Its application to the Blazhko star MW Lyr (Jurcsik et al. 2009) shows how the mean global parameters, such as the radius, luminosity and surface effective temperature of a modulated star vary over the Blazhko cycle, a very important result for the further development of Blazhko models.

The magnetic model for the Blazhko effect received some blows when spectropolarimetric observations indicated that a strong kiloGauss-order dipole field, a premise in the model (Shibahashi 2000), is not present in the prototype RR Lyr (Chadid et al, 2004) nor in a sample of 17 selected modulated and non-modulated RR Lyrae stars (Kolenberg \& Bagnulo 2009).

\begin{figure*}
\centering
%\includegraphics[width=12cm]{mgsio2.eps}
%\plotone{Kolenberg_Figure4.eps}
\plotone{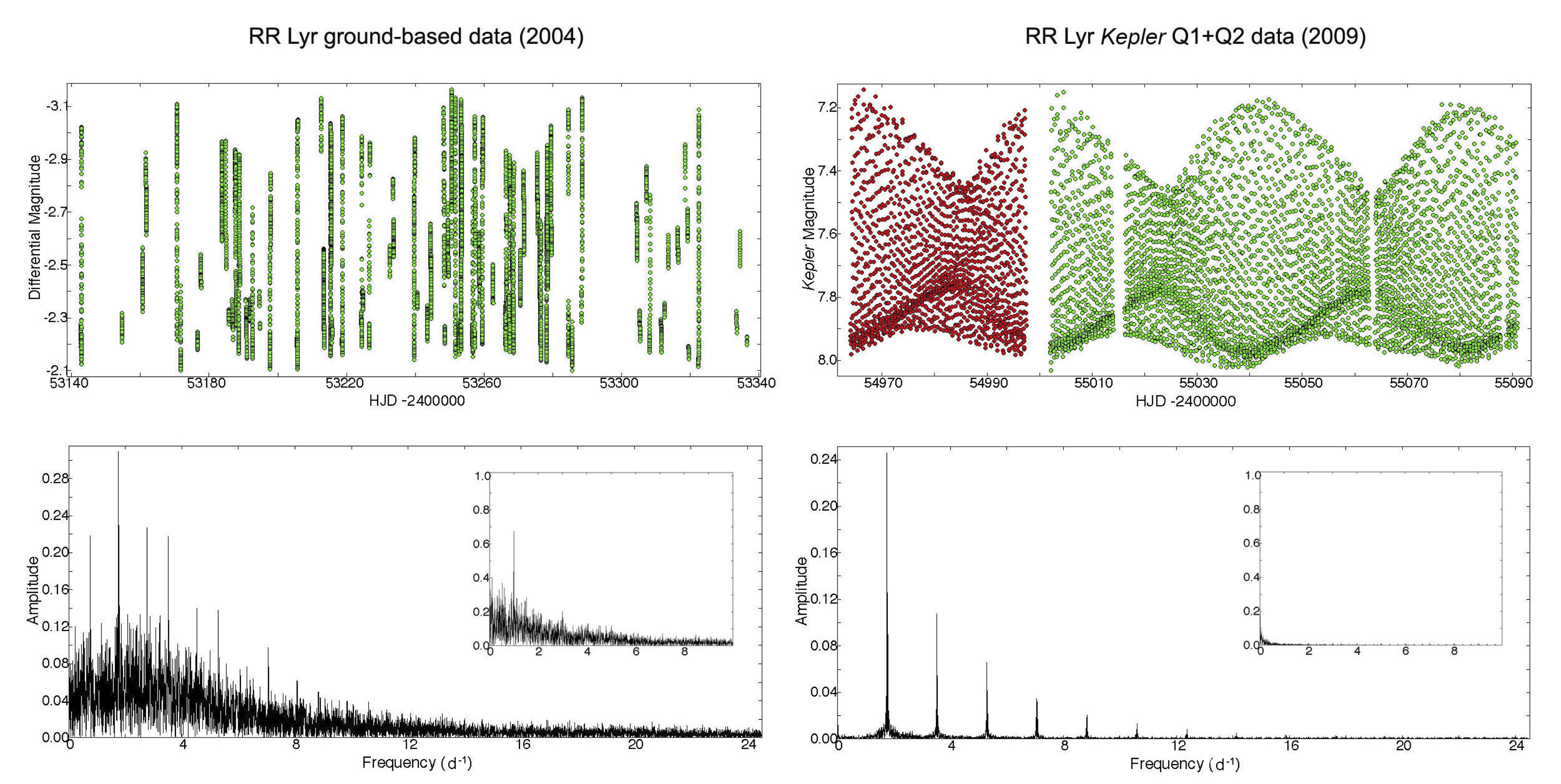}
\vskip0pt
\caption{ Comparison of the ground-based RR~Lyr data, gathered from six different observatories, published by Kolenberg et al. (2006) and the {\it Kepler} Q1 and Q2 data of the star transformed to the magnitude scale (top panels).  A folded light curve of the Q1 data is shown in Figure\,1 (lower right panel). Bottom panels: Fourier transform of the data; the insert shows the window function. Figure from Kolenberg et al. (2011a).}
\label{mgsio2}
\end{figure*}

Several space telescopes have asteroseismology as an important part (or prime by-product) of their mission and have performed unprecedented observations of RR Lyrae stars:  the Canadian suitcase-sized {\it MOST} telescope (Gruberbauer et al. 2007), the French-led ESA space mission {\it CoRoT} (Chadid et al. 2010), and NASA's {\it Kepler} mission (Kolenberg et al. 2010a). The huge advantage of stellar data gathered by space missions is illustrated in Figure 4 (from Kolenberg et al. 2011).  This figure shows the quality of the Kepler data of RR Lyr compared to one of the most precise published data sets on the star obtained from ground-based observatories (Kolenberg et al. 2006). Despite the effort of organizing a multi-site campaign and combining all the standardized observations, the latter data set had a point-to-point scatter of over 0.005 mag at best, and was characterized by nightly and weather gaps, typical for Earth-based observations.
As a consequence, the Fourier spectra of such data set obtained from the ground have a noise level 10-50 times higher than Kepler Fourier spectra and are subject to aliasing.  

For Blazhko stars, data from both the {\it CoRoT} and {\it Kepler} space missions have revealed complex multiplet structures in the Fourier spectra and additional frequency peaks of which some can be explained in terms of radial overtones, and others not (Chadid et al. 2010, Benk\H{o} et al. 2010).  

An immediate result obtained from the new {\it CoRoT} data (Szab\'o et al. 2010a) and {\it Kepler} data (Benk\H{o} et al. 2010) is that half of the RR Lyrae stars observed by {\it CoRoT} and {\it Kepler} show Blazhko modulation, confirming the recent findings from the dedicated ground-based Konkoly survey (Jurcsik et al. 2010) whereas older ground-based studies mentioned only 20-30\% (Szeidl 1988).  This discrepancy is mostly due the precision and duty cycle of the recent observations. The {\it Kepler} data made it possible to find a Blazhko modulation with an amplitude as small as 0.03 mag, and a period (phase) modulation of $\delta P_0 < 1.5$\,min, the smallest modulation values known so far (Benk\H{o} et al. 2011).
With increasing occurrence rates for the Blazhko effect, one could ask whether maybe all RR Lyrae stars show the Blahko effect to some degree. However, even with Kepler precision, several RR Lyrae stars turn out to be unmodulated.  Interesting results on the non-Blazhko stars in the Kepler sample were obtained by Nemec et al. (2011, and this proceedings).

\section{Serendipity and Progress}

The above-mentioned results contributed to a considerable progress in our understanding of the Blazhko effect.  But a major breakthrough resulted from serendipity.  
By a fortunate coincidence, the star RR Lyr, the brightest star and prototype of its class and a Blazhko variable studied since over a century, is located in the Kepler field. 
The star has a visual apparent magnitude between 7.2 and 8.2 and hence it was initially thought to be too bright to be observed successfully with Kepler. 
The first {\it Kepler} findings described in Kolenberg et al. (2010a, see below), based on the first 34 days of Kepler observations of the star, however, triggered further interest in the star.  Therefore, a custom aperture was devised for RR~Lyr, reducing the amount of originally assigned pixels from 433 to about 150.  Thanks to this calibration work by Steve Bryson (NASA Ames) described by Kolenberg et al. (2011), RR Lyr could be scheduled as a {\it Kepler} target. Over the past year, it has been observed through the {\it Kepler} Guest Observer (GO) program {\rm (http://keplergo.arc.nasa.gov/)} and the {\it Kepler} Asteroseismic Science Consortium (KASC, http://astro.phys.au.dk/KASC/). 
In this way, RR Lyr became a prime example of a bright (and famous) star observed with the {\it Kepler} satellite.  The general philosophy is not to waste any pixels on a star of which one cannot capture all the useful flux.  
But RR Lyr is an unusual case due to its high variability. There are measurements in which the star is well captured and there is a lot of saturated flux in columns captured in every measurement, thus one can make good estimates of missing flux and recover photometry.  Such unusual cases were not considered earlier in the {\it Kepler} Mission.

\begin{figure*}
\centering
\includegraphics[width=9cm]{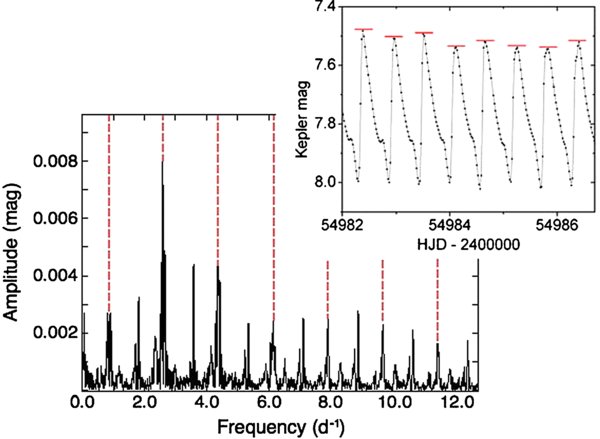}
%\plotone{PD1.jpg}
\vskip0pt
\caption{Residual spectrum of RR Lyr after subtraction of the main frequency $f_0$, its harmonics, and the triplet components. The dashed lines mark the positions of the half-integer frequency components. The highest peak occurs at $\pm 3/2f_0$. The inset shows the alternating heights of the maxima for the star, connected to the additional frequencies (from Kolenberg et al 2010a).}
\label{PD1}
\end{figure*}

Already in the first quarter (33.5 days) of released {\it Kepler} data of the star we detected a new phenomenon, reflected in alternatingly high and low maxima (Kolenberg et al. 2010a). In the frequency spectrum, these variations with the double period result in the appearance of so-called haf-integer frequencies (see Figure\,\ref{PD1}). This phenomenon, called period doubling, was theoretically predicted in Cepheids (see, e.g., Moskalik \& Buchler, 1990, ApJ 355, 590), stars that undergo radial pulsations just like RR Lyrae stars, but had never been observed so far, in Cepheids nor in RR Lyrae stars. Within the KASC RR Lyrae working group, the phenomenon was studied in more detail, see, e.g., Kolenberg et al. (2011) and Szab\'o et al. (2010b).  In the latter paper, the phenomenon is investigated in detail and the authors find that a 9:2 resonance between the fundamental radial mode and the 9th overtone might be responsible for period doubling.  The fact that period doubling was also found in a few other Blazhko stars, and not in non-modulated stars, revealed a possible connection between period doubling and modulation. 

\begin{figure*}
\centering
%\includegraphics[width=12cm]{mgsio2.eps}
%\plotone{Kolenberg_Figure6.eps}
\plotone{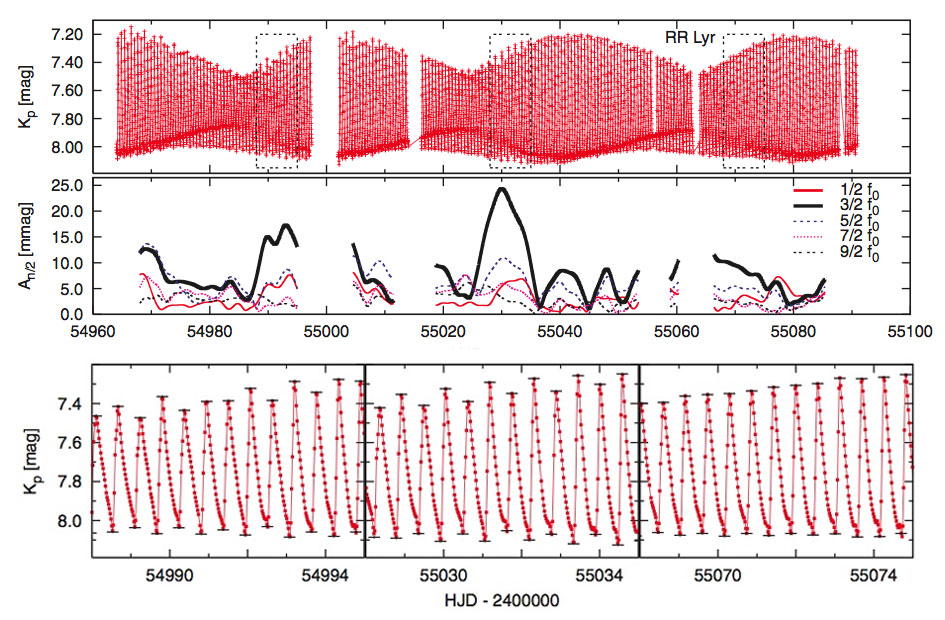}
\vskip0pt
\caption{Light curve of RR Lyr from Q1 and Q2. Note that the individual pulsational cycles are hardly discernible. The (39.6-d) Blazhko modulation clearly stands out. The three dashed boxes are enlarged in the bottom panel. They highlight the same phase interval in the Blazhko cycle for consecutive modulation cycles.  The occurrence of the period doubling effect is strongly variable. The middle panel shows the amplitudes of the half-integer frequencies (from Szab\'o et al. 2010b).}
\label{PD2}
\end{figure*}

In the subset of Blazhko stars that show period doubling, we observe that it does not occur at all phases in the Blazhko cycle.  It can be very obvious in some phases, and invisible in others (see Figure\,\ref{PD2}).
There is also strong evidence that the Blazhko modulation is strongly variable in several stars, where our data clearly indicate that there is no strict repetition from one Blazhko cycle to the next.  Also period doubling does not repeat in the same Blazhko phases for consecutive cycles (Szab\'o et al. 2010b).

Why was period doubling, which in some phases in the Blazhko cycle reaches a rather high amplitude, not detected earlier in ground-based data, particularly for a star as well-studied as RR Lyr, the prototype of the class?  Besides the fact that period doubling does not always occur, the main reason is undoubtedly that RR Lyrae stars have periods of around half a day, and, when observing from one site on Earth, consecutive pulsation maxima are usually missed.

\begin{figure*}
\centering
%\includegraphics[width=12cm]{mgsio2.eps}
%\plotone{Kolenberg_Figure7.eps}
\plotone{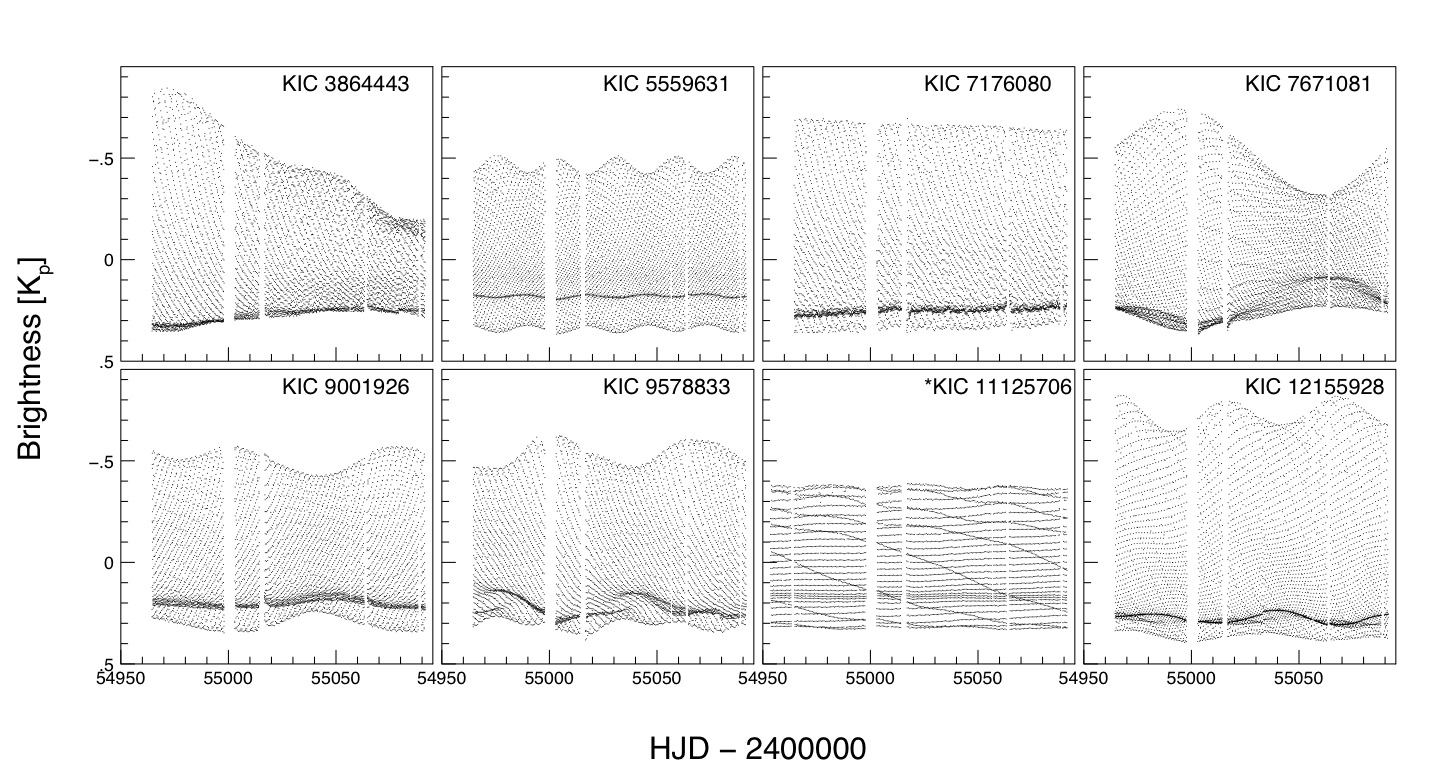}
\vskip0pt
\caption{A glimpse of Kepler's Blazhko zoo, based on data from Kepler quarters Q1 and Q2 (from Benk\H{o} et al. 2010).     }
\label{o1039}
\end{figure*}

\section{A new model hierarchy}

The observation of period doubling sparked new modelling efforts (Szab\'o et al. 2010b, Koll\'ath et al. 2011), and recently a new model for the Blazhko effect was proposed by Buchler \& Koll\'ath (2011).  Using the amplitude equation formalism to study the nonlinear, resonant interaction between the two modes, they showed that the (9:2) radial resonance can not only cause period doubling, but it may also lead to amplitude modulation. Moreover, in a broad range of parameters the modulations can be irregular, just like recent observations show. 

At the same time, the possibility of the Stothers (2006, 2010) scenario was investigated in more detail by Moln\'ar \& Koll\'ath (2010) and Smolec et al. (2011). The latter authors used the new Kepler data for a very detailed comparison with theoretical models.
They tested the Stothers model by periodic variation of the turbulent
convection (mixing length) in hydrodynamic models. The variation of the light curve is described in terms of Fourier
decomposition parameters and compared with the observations.
They found that, in order to get the same ranges of variation of the Fourier parameters as are
observed in RR Lyr, a modulation of the mixing length parameter $\alpha$ on the order of 50 per cent on 
a short time scale is needed, which is a very large, possibly physically implausible variation.
In addition, Smolec et al. (2011) were able to reproduce the period doubling phenomenon in their models.

% ADD ANIMATION LINK ? 

\begin{figure*}
\centering
%\includegraphics[width=12cm]{mgsio2.eps}
%\plotone{Kolenberg_Figure8.eps}
\plotone{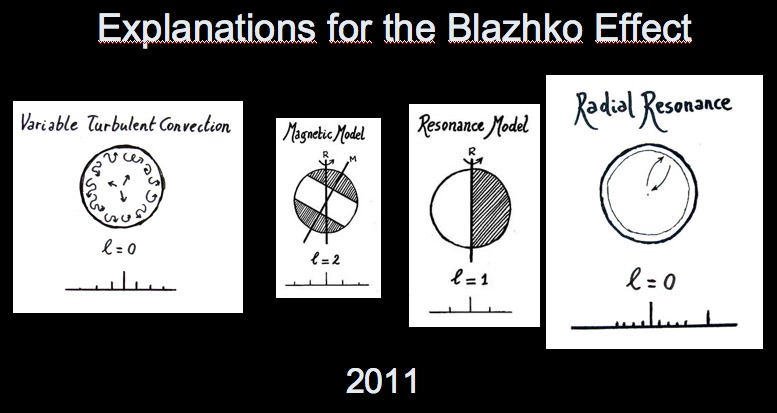}
\vskip0pt
\caption{A new hierarchy for the models for explaining the Blazhko effect.}
\label{expl2}
\end{figure*}

Therefore, at this point in time,  we are left with a different pictures and a new hierarchy for the models for explaining the Blazhko effect (Figure\,\ref{expl2}).

\newpage

\section{A new era}

The new, unprecedentedly precise data delivered by the {\it CoRoT} and {\it Kepler} mission (see Figure\,7, see also Karen Kinemuchi's contribution, this proceedings) of RR Lyrae stars reveal previously unseen features in Blazhko stars, such as period doubling, and also additional frequencies, which we do not fully understand yet.
A few fundamental-mode pulsators show variations with an additional period that could be the second overtone (Poretti et al. 2010, Benk\H{o} et al. 2010, Nemec et al. 2011).  Other frequency peaks occur at positions where their period ratio cannot be explained in terms of radial modes, as we know them.

Over the past past decades, there have been several observations that hint at the presence of nonradial modes in RR Lyrae stars (e.g., Olech et al. 2000).
However, additional peaks that cannot be identified with radial modes and, for the specific case of Blazhko stars, components in a multiplet, do not necessarily imply that the frequencies must be related to a nonradial mode.  
As pointed out by Kov\'acs (2009), accurate time-series spectral line analysis is needed to reveal any possible non-radial components.  Only this would definitively allow us to include (or exclude) non-radial modes in explaining the Blazhko phenomenon. 

Benk\H{o} et al. (2011) present an analytical formalism for the description of Blazhko RR Lyrae light curves in which they employ a treatment for the amplitude and frequency modulations in a manner similar to the theory of electronic signal transmitting. Earlier work treating Blazhko RR Lyrae light curves as modulated signals was done by Szeidl \& Jurcsik (2009) and Benk\H{o} et al. (2010).  
The method naturally explains numerous light curve characteristics and explains properties of the Fourier spectra.  In addition, this formalism significantly reduces the number of free parameters in the light curve solutions.  Therefore, an analysis of the available light curve data in this framework may reveal additional constraints that help us to better understand the Blazhko phenomenon.

Approaches to derive atmospheric parameters of RR Lyrae stars through multicolor spectroscopy and/or spectroscopy, such as applied by De Boer \& Maintz (2010), S\'odor et al. (2009), Kolenberg et al. (2010b) and For et al. (2011), allow us to connect the complex atmospheric variations with the pulsation patterns, and even their modulations.  In addition, new and exciting findings such as the occurrence of Helium emission (Preston et al. 2009) and neutral line disappearance (Chadid et al. 2009) would be very interesting to study as a function of the Blazhko phase. 
These methods, and their extensions, in combination with the information from the high-precision, quasi-uninterrupted data delivered by the satellite missions, provide (literally!) in-depth studies of the Blazhko phenomenon.

\section{Acknowledgements}
George, your curiosity and enthusiasm over all those decennia has largely contributed to the treasure chest we are currently examining, and your kindness and "joie de vivre" encourages many to follow your example.  Thank You, and Happy Birthday!   

I thank Andy McWilliam for inviting me to present at this exceptional, historic and stimulating meeting and for making it a success, and for his organizational skills and patience! 
The entire Kepler Team and KASC (Kepler Asteroseismic Science Consortium) is greatly acknowledged for its outstanding efforts that have made recent breakthroughs in astrophysics possible.  Particular thanks go to Dr. Steve Bryson (NASA Ames) who invested time and energy in devising a custom aperture for RR Lyrae. Finally, the members of the KASC RR Lyrae working group are acknowledged for their enthusiasm and dedicated work.  Jozsef Benk\H{o} and Robert Szab\'o are kindly thanked for Figures 6 and 7.

I am grateful for the financial support of Austrian FWF project  P 21830-N16 that allowed me to attend the meeting, and am currently a Marie Curie Fellow under FP7 grant 255267 between the University of Leuven (Belgium) and Harvard University (US).

\end{document}